\documentclass[%
 aip,
 apl,%
 amsmath,amssymb,
 reprint,%
]{revtex4-1}

\usepackage{graphicx}
\usepackage{dcolumn}
\usepackage{bm}
\usepackage{color}

\begin{document}


\title[]{Versatile Strain-Tuning of Modulated Long-Period Magnetic Structures}

\author{D. M. Fobes}
\author{Yongkang Luo}
\author{N. Le\'{o}n-Brito}
\author{E. D. Bauer}
\affiliation{MPA-CMMS, Los Alamos National Laboratory, Los Alamos, NM 87545, USA}%
\author{V. R. Fanelli}
\affiliation{P-27, Los Alamos National Laboratory, Los Alamos, NM 87545, USA}%
\affiliation{ISD, Oak Ridge National Laboratory, Oak Ridge, TN 37831, USA}%
\author{M. A. Taylor}
\affiliation{P-27, Los Alamos National Laboratory, Los Alamos, NM 87545, USA}%
\author{L. M. DeBeer-Schmitt}
\affiliation{CEMD, Oak Ridge National Laboratory, Oak Ridge, TN 37831, USA}%
\author{M. Janoschek}
\affiliation{MPA-CMMS, Los Alamos National Laboratory, Los Alamos, NM 87545, USA}%

\date{\today}

\begin{abstract}
We report a detailed small-angle neutron scattering (SANS) study of the skyrmion lattice phase of MnSi under compressive and tensile strain. In particular, we demonstrate that tensile strain applied in the skyrmion lattice plane, perpendicular to the magnetic field, acts to destabilize the skyrmion lattice phase. This experiment was enabled by our development of a  versatile strain cell, unique in its ability to select the application of either tensile or compressive strain \textit{in-situ} by using two independent helium-actuated copper pressure transducers, whose design has been optimized for magnetic SANS on modulated long-period magnetic structures and vortex lattices, and is compact enough to fit in common sample environments, such as cryostats and superconducting magnets.
\end{abstract}

\maketitle

Modulated long-period magnetic structures incommensurate to the underlying crystal structure have been demonstrated to be at the heart of a wide range of phenomena in solid-state physics such as the magneto-electric coupling in multiferroic materials,\cite{Cheong07} the magneto-caloric effect,\cite{Tishin2003}, the emergence of topologically-stabilized magnetic defects such as solitons\cite{Janoschek2010} and skyrmions,\cite{Muehlbauer2009} and unconventional superconductivity. In the latter case, long-period magnetic arrangements arise in two distinct forms, namely the underlying long-range ordered antiferromagnetic state whose critical fluctuations are believed to mediate unconventional superconductivity,\cite{Scalapino2012} and the vortex lattice that forms in type-II superconductors,\cite{Eskildsen2011} which in some cases are intimately coupled to one another.\cite{Eskildsen1998} This makes long-period magnetic structures relevant for applications ranging from solid-state cooling to energy-applications to data storage and spintronics.

Strain frequently represents an important tuning parameter in the optimization of the desired material response related to these phenomena. For example, in multiferroic materials, the ferroelectric polarization is strongly coupled to the underlying crystal structure, which can be tuned by strain, and will affect the magnetization via magnetostriction.\cite{Windsor2014} Likewise, uniaxial strain can be utilized to alter magnetic exchange integral overlap along a specific direction, and thus introduce or enhance magnetic anisotropy. It has also been demonstrated that magnetic anisotropy in magnetocaloric materials results in the rotating magnetocaloric effect, which is believed to facilitate the implementation of magnetic cooling.\cite{Nikitin2010} Additionally, uniaxial magnetic anisotropy has been shown to increase the stability of skyrmion lattice phases.\cite{Banerjee2014, Nii2015, Shibata2015, Lin2015, Chacon2015} Finally, in the iron pnictide family of materials the antiferromagnetic and unconventional superconducting states show extreme sensitivity to strain.\cite{Chu2012}

Small-angle neutron scattering (SANS) has proven itself to be one of the most powerful probes to study relevant microscopic parameters of modulated long-period magnetic structures, such as the order parameter and the period, and is a crucial tool for the study of superconducting vortex lattices.\cite{Eskildsen2011} However, most available \textit{in-situ} strain application options for SANS beam lines typically consist of large load frames designed for mechanical strain testing of micro-structures. They are therefore large installations allowing for neither cooling of the sample below room temperature, nor application of magnetic fields, both required for the investigation of modulated long-period magnetic structures. Small strain cells that enable investigation of the effects of strain on electrical transport, bulk magnetization, or nuclear magnetic resonance at low temperatures have been developed,\cite{Hicks2014} and can be obtained commercially in some cases, but, due to the relatively large samples ($\gtrsim 1$~mm$^3$) required for SANS, these strain cells are unsuitable for this purpose.

\begin{figure}[!htbp]
\vspace{-0.25cm}
\begin{center}
    \includegraphics[width=1\columnwidth]{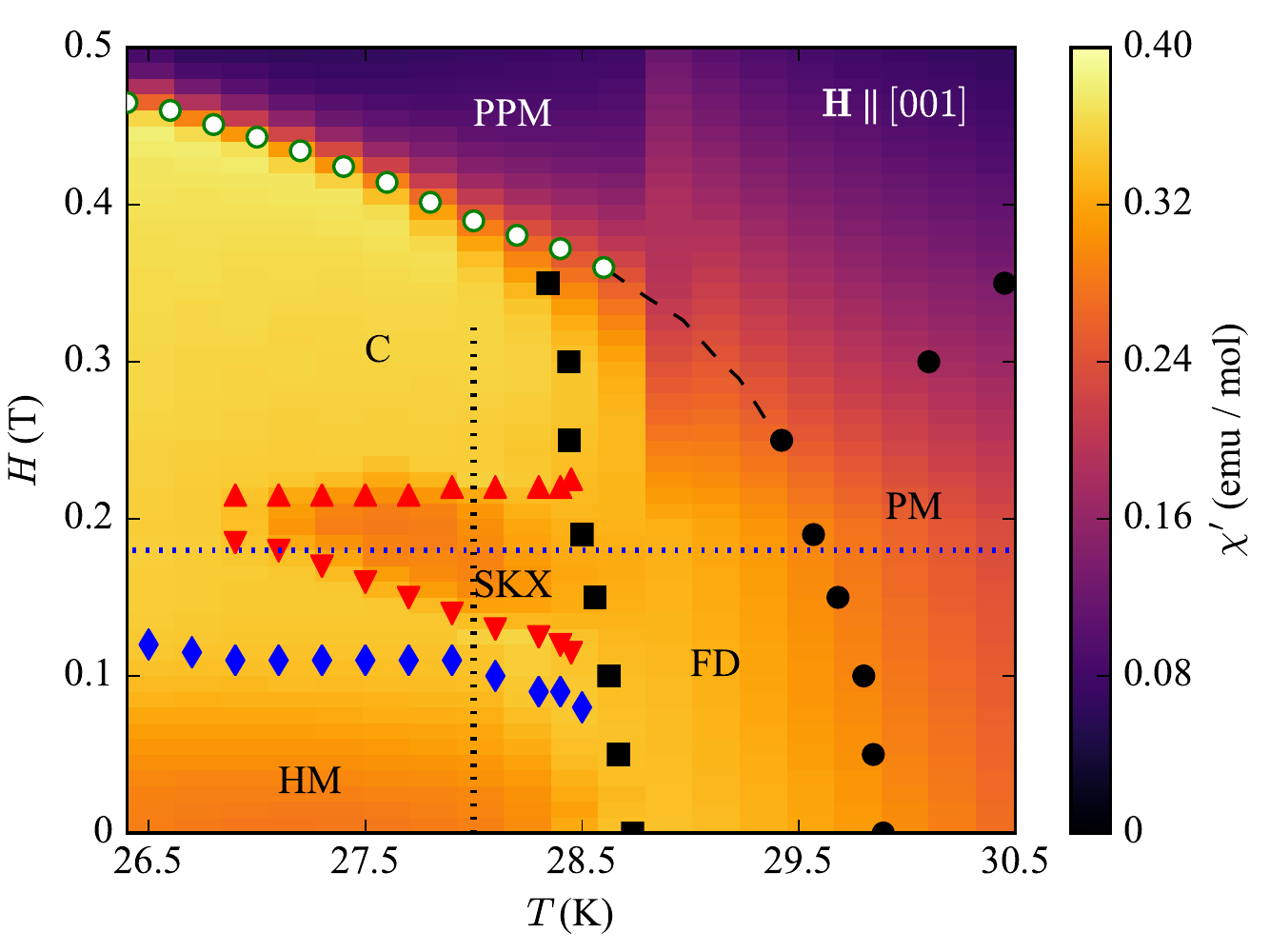}
    \vspace{-0.5cm}
    \caption{Phase diagram of MnSi, where the color scale represents the magnitude of the ac magnetic susceptibility $\chi^{\prime}$. PM: paramagnetic phase, FD: fluctuation disordered phase, HM: helimagnetic phase, C: conical phase, SKX: skyrmion, or A-phase, PPM: field polarized paramagnetic phase. The magnetic field was aligned along [001]. Green circle, red upwards and downwards triangle, and blue diamond symbols were extracted from the shown ac susceptibility. Black squares and circles were obtained from resonant ultrasound spectroscopy (RUS) measurements in Ref. \onlinecite{Yongkang2017}, performed on a small piece of the same sample. Dashed horizontal and vertical lines represent measurement trajectories for data used for Fig. \ref{fig:SANSdata}.}
    \label{fig:phase diagram}
\end{center}
\vspace{-0.75cm}
\end{figure}

Here we report the effects of strain on the stability of a skyrmion lattice, whose investigation motivated the  development of a strain cell specifically for use on a SANS beam line, and intended as a general tool for the investigation of the influence of strain-tuning on modulated long-period magnetic structures. Using our strain cell design we have investigated the strain response of the skyrmion lattice in MnSi, the material first reported to exhibit a magnetic skyrmion lattice. This choice offers the advantage that MnSi is well-characterized both experimentally and theoretically.\cite{Nagaosa2013} In the class of cubic B20 materials to which MnSi belongs, competition between ferromagnetic exchange $J$ and the Dzyaloshinskii-Moriya $D$ interaction results in long-period helimagnetic ordering below a temperature $T_c$ ($T_c = 29$~K for MnSi).\cite{Bak1980} Application of a small magnetic field $H$ stabilizes a skyrmion lattice in the plane $\perp H$ in a small phase pocket below $T_c$ (historically known as the A-phase, cf. Fig. \ref{fig:phase diagram}). The skyrmion lattice parameter is proportional to $D/J$, $\lambda = 18$~nm for MnSi.\cite{Muehlbauer2009} These topologically-protected magnetic textures are only weakly coupled to the crystal structure, enabling skyrmions to move independently from the underlying crystal. Notably, ultra-low current densities of 10$^6$~A/m$^2$ are sufficient to move skyrmions via the spin-torque effect,\cite{Schulz2012} which makes skyrmions promising for memory and spintronics applications.\cite{Fert2013, Nagaosa2013}

The interest in the strain response of skyrmions is two-fold: (a) on one hand, many future devices will be designed from thin films and nanostructures,\cite{Iwasaki2013, Jiang2015} which are typically under epitaxial strain. (b) On the other, both theory\cite{Butenko2010, Banerjee2014, Lin2015} and experiment \cite{Nii2015, Shibata2015, Chacon2015} have shown that in the class of cubic B20 materials, the application of magnetic anisotropy $A$ via strain controls the stability of the skyrmion phase. For example, it has been demonstrated that compressive uniaxial pressure $\sigma$ applied in the plane perpendicular to $H$ acts as uniaxial magnetic anisotropy and increases the tiny size of the A-phase.\cite{Nii2015, Chacon2015} It was recently suggested that a significant enough easy-axis anisotropy along the strain axis is generated regardless of the crystallographic orientation, although in that study strain was only applied along the three principle cubic symmetry axis [100], [110] and [111].\cite{Chacon2015} Considering that the size of the A-phase in MnSi changes dramatically by merely applying  $H$ along the [100] or [111] directions (for $\sigma=0$),\cite{Bauer2012} this is an open question.  It was also shown that for $\sigma\parallel H$ the skyrmion lattice is destabilized, but the response is weaker by nearly a factor two.\cite{Nii2015, Chacon2015}

In stark contrast, the response of the skyrmion phase to tensile stress remains experimentally unexplored. Mean-field calculations \cite{Butenko2010} predict that tensile stress parallel to $H$ should stabilize the skyrmion lattice, which suggests, by analogue to the previous compressive stress experimental results, that tensile strain perpendicular to $H$ will decrease the size of the A-phase. However, because the magnitude of the response observed for compression differs significantly for $\sigma\perp H$ and $\sigma\parallel H$, this raises the question whether the response differs as function of the direction of the strain, or whether the response for stabilizing/destabilizing the A-phase is generically different.

To answer these open questions, we designed a strain cell, described in detail below, keeping in mind the following key requirements: (i) the cell should be compact enough to fit in standard sample environment options such as low-temperature cryostats and superconducting magnets, typically available at SANS instruments;  (ii) it should allow continuous changes in strain \textit{in-situ}; (iii) and most importantly, as an unique advantage over strain cells previously developed for magnetic SANS\cite{Chacon2015} it should allow for the application of both tensile and compressive strain. We note that the latter requirement is not only a key feature required to make progress on the issues in skyrmion physics described above but is generally advantageous when investigating the influence of strain on modulated long-period magnetic structures. For example, in most cases it is not clear whether tension or compression will induce the desired type of magnetic anisotropy (for example easy-axis vs. easy-plane). Further, in case the investigated material exhibits a hysteretic response, it is often helpful to record full strain-loops (from positive to negative strain, corresponding to compression and tension). As we will demonstrate below, our strain cell meets these key requirements.

The single crystal of MnSi used for our experiments was grown via the Bridgman method, and characterized via x-ray Laue diffraction. Electrical resistivity measurements carried out in a Quantum Design Physical Property Measurement System (PPMS) with the current applied along the [100] direction on a small piece of this sample reveal a residual resistivity ratio, defined as the ratio of electrical resistivity at 300~K and 2~K, of 87, comparable to samples used in previous neutron scattering studies on MnSi.\cite{Muehlbauer2009, Janoschek2010a, Janoschek2013} ac magnetic susceptibility measurements shown in Fig. \ref{fig:phase diagram} were performed in a Quantum Design Magnetic Property Measurement System (MPMS). These measurements determine the boundaries of the skyrmion-lattice phase in Fig. \ref{fig:phase diagram} via pronounced peaks in the real part of ac susceptibility. SANS measurements were carried out at the GP-SANS instrument at Oak Ridge National Laboratory's High Flux Isotope Reactor, using a neutron wave length of $\lambda = 6$~\AA\ ($\Delta \lambda / \lambda = 15 \%$) and 10.75~m sample-detector distance. Magnetic field was applied along the incident beam using an 11~T superconducting magnet. 

\begin{figure}[!htbp]
\vspace{-0cm}
\begin{center}
    \includegraphics[width=0.85\columnwidth]{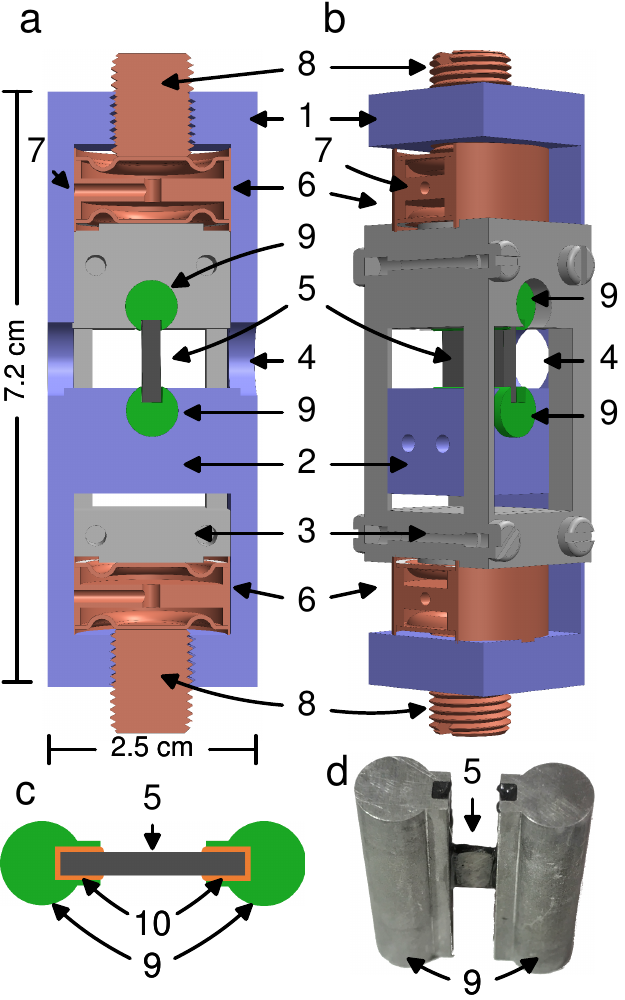}
    \vspace{-0.5cm}
    \caption{Schematic of the strain cell, (a) cut parallel to neutron beam, through the center of the device, and (b) cut perpendicular to neutron beam, through primary load frame. (c) Schematic of epoxy-mounted sample in brackets. (d) Photo of the MnSi plate-like sample epoxy-mounted in brackets. (1) Primary load frame, (2) cross-beam fixed to load frame, (3) floating elevator, (4) neutron beam apertures, (5) sample, (6) copper pressure transducers, (7) He gas lines, (8) pre-tension screws, (9) sample brackets, (10) Stycast 2850FT epoxy.}
    \label{fig:strain cell}
\end{center}
\vspace{-1cm}
\end{figure}

The strain cell used and developed for these experiments is shown in Fig. \ref{fig:strain cell}. We note that strain cell designs for bulk measurements, which expect significantly smaller samples, utilize piezo-electric stacks to apply strain,\cite{Hicks2014} the advantages being their ability to produce large forces on the sample, and easy controllability via applied voltage. However, the stroke of piezo-electric stacks is only in the range of a few $\mu$m, and therefore cannot be utilized for the large samples required for SANS (typically $> 1$~mm$^3$). To overcome this limitation, our design uses copper pressure transducers, developed in-house at Los Alamos National Laboratory, consisting of cylindrical copper bodies containing empty volumes at the top and the bottom, over which pre-indented copper bellows are soldered. Helium gas pressure, controlled and maintained via a GE Druck PACE 5000 pressure controller, is used to actuate the bellows. Finite Element Analysis performed in Solid Works determined that a He gas pressure of 1200~psi will deflect the copper bellows by 75~$\mu$m and generate a maximum force $F_{\rm max} = 270$~N onto the sample, which was confirmed using a well-defined copper sample with an attached strain gauge. The 1200~psi maximum helium gas pressure is defined by the solder joints, and the pressure transducer gas lines are heated at low temperatures to prevent solidification. We note that for a bar-like sample with a size of $1 \times 1 \times 10$~mm$^3$, the maximum force of 270~N corresponds to an uniaxial pressure of $\sigma_{\rm max} = 270$~MPa. Because typical metallic samples  have bulk moduli $E$ of approximately 100~GPa, in principle this should achieve strains of approximately $\varepsilon=\sigma/E = 0.27$\%, comparable to most strain experiments. The real strain is, of course, slightly smaller due to transfer of some applied pressure into the load frame.

The strain cell body was designed to fit sample environments commonly available for magnetic SANS, such as low temperature cryostats and superconducting magnets, limiting its diameter to $<40$~mm. Hardened 7075 aluminum was chosen as a construction material due to its lower background in neutron scatteringa resulting from a low coherent scattering cross-section, and large mechanical strength. To allow for the application of strain in both compressive and tensile modes \textit{in-situ}, the strain cell consists of several independent parts: (i) a primary load frame (1) with a fixed cross-beam (2) and (ii) a floating elevator (3). The sample is affixed to the fixed cross-beam and to the elevator, allowing vertical movement within the primary load frame. Two \textit{independent} copper bellow pressure transducers (6) mounted above and below the elevator move the elevator up or down within the primary load frame, extending or compressing the sample, respectively. To simplify sample changes, we developed a system where the sample (5) is attached to cylindrical aluminum brackets, enabling effortless sliding of the sample receptacles into the load frame and elevator, (9) using Stycast 2850FT epoxy (10). The brackets ensure that the sample is fully embedded in epoxy at its attachment points; this method has the advantage of a more even strain distribution that prevents "bowing" of the sample which was observed in a previously report where the sample was attached using on one of its long surfaces.\cite{Hicks2014}. Stycast epoxy was used due to its known ability to withstand high pressures, relevant for compression,\cite{Walker1999} but equally due to its demonstrated favorable cryogenic tensile properties. Notably, van de Camp \textit{et al.} determined that the tensile strength of Al 2024 / Stycast 2850FT composite sandwiches, highly similar to our brackets, doubles when cooled from room temperature to 77 K, at which it is $\sim100$~MPa.\cite{Camp2015} Furthermore, for Stycast 1266, the unfilled variant of Stycast 2850FT, the tensile fracture strain is known to be 4\% \cite{Hicks2014}, well above the strain values achievable with our cell design. The neutron beam enters and exits the strain cell through two 8.9~mm diameter apertures (4), allowing for a maximum scattering angle of $\pm20^{\circ}$, more than large enough for typical SANS measurements. A small Cd aperture (3~mm) was affixed to the entry aperture to reduce background.

\begin{figure}[!t]
\begin{center}
    \includegraphics[width=0.9\columnwidth]{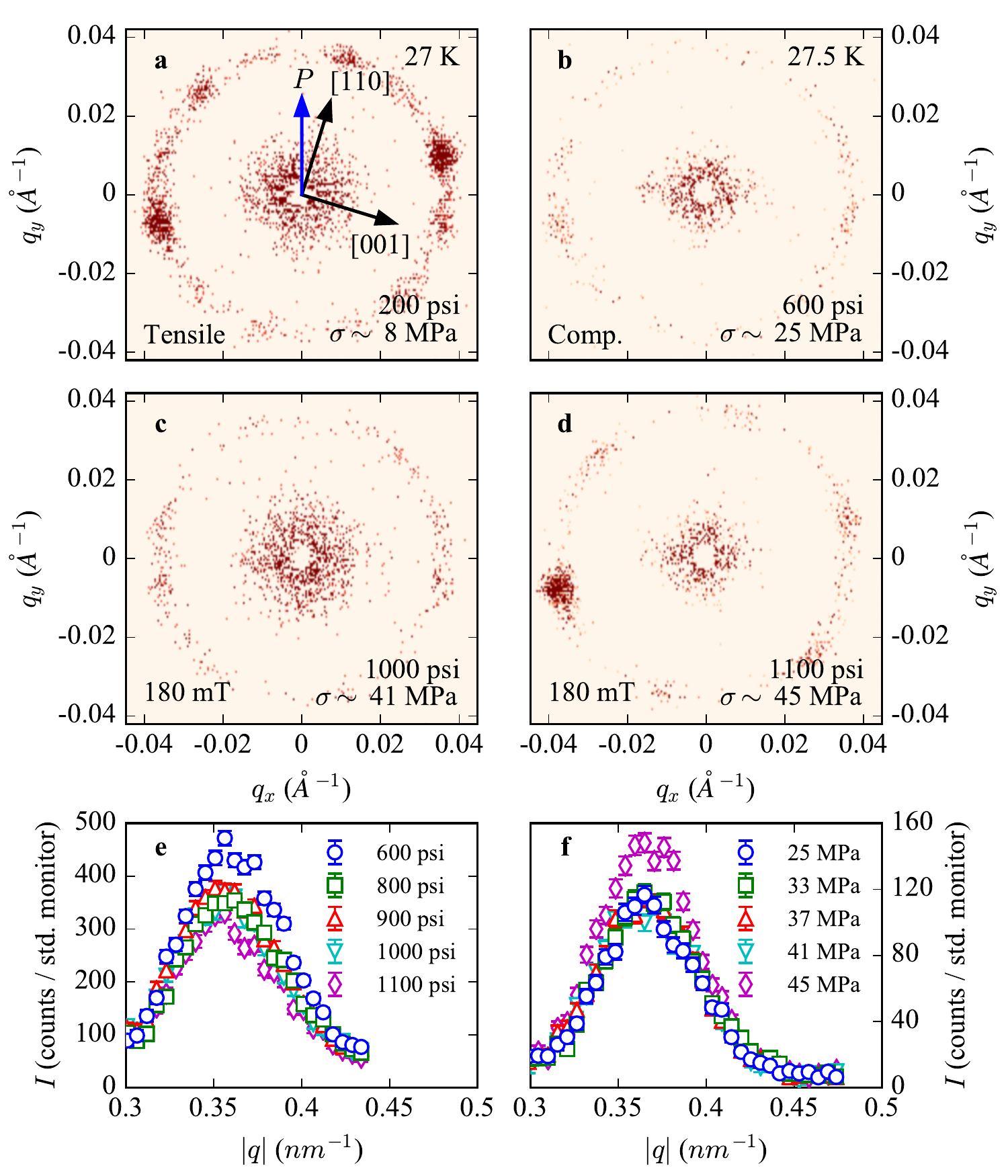}
    \vspace{-0.25cm}
    \caption{SANS detector images under tensile (left column) and compressive (right column) strain. Detector images under tensile strain shown at (a) 200~psi ($\sigma \approx 8$~MPa) and (c) 1000~psi (41~MPa), and compressive strain at (b) 600~psi (25~MPa) and (d) 1100~psi (45~MPa). All data are obtained at 27~K and 0.18~T, near the edge of the A-phase. q-averaged integrated intensity for (e) tensile and (f) compressive strain for 600 (blue circles), 800 (green squares), 900 (red upwards triangles), 1000 (cyan downwards triangles), and 1100~psi He gas pressure (magenta diamonds). Approximate equivalent uniaxial pressures are shown in the legend in (f). Strain is applied vertically (line labeled $P$), and the sample is oriented with $[1\bar{1}0] \parallel $ beam, misaligned clockwise by $17^{\circ}$, shown in (a).}
    \label{fig:detector}
\end{center}
\vspace{-1cm}
\end{figure}

\begin{figure}[!htbp]
\begin{center}
    \includegraphics[width=0.9\columnwidth]{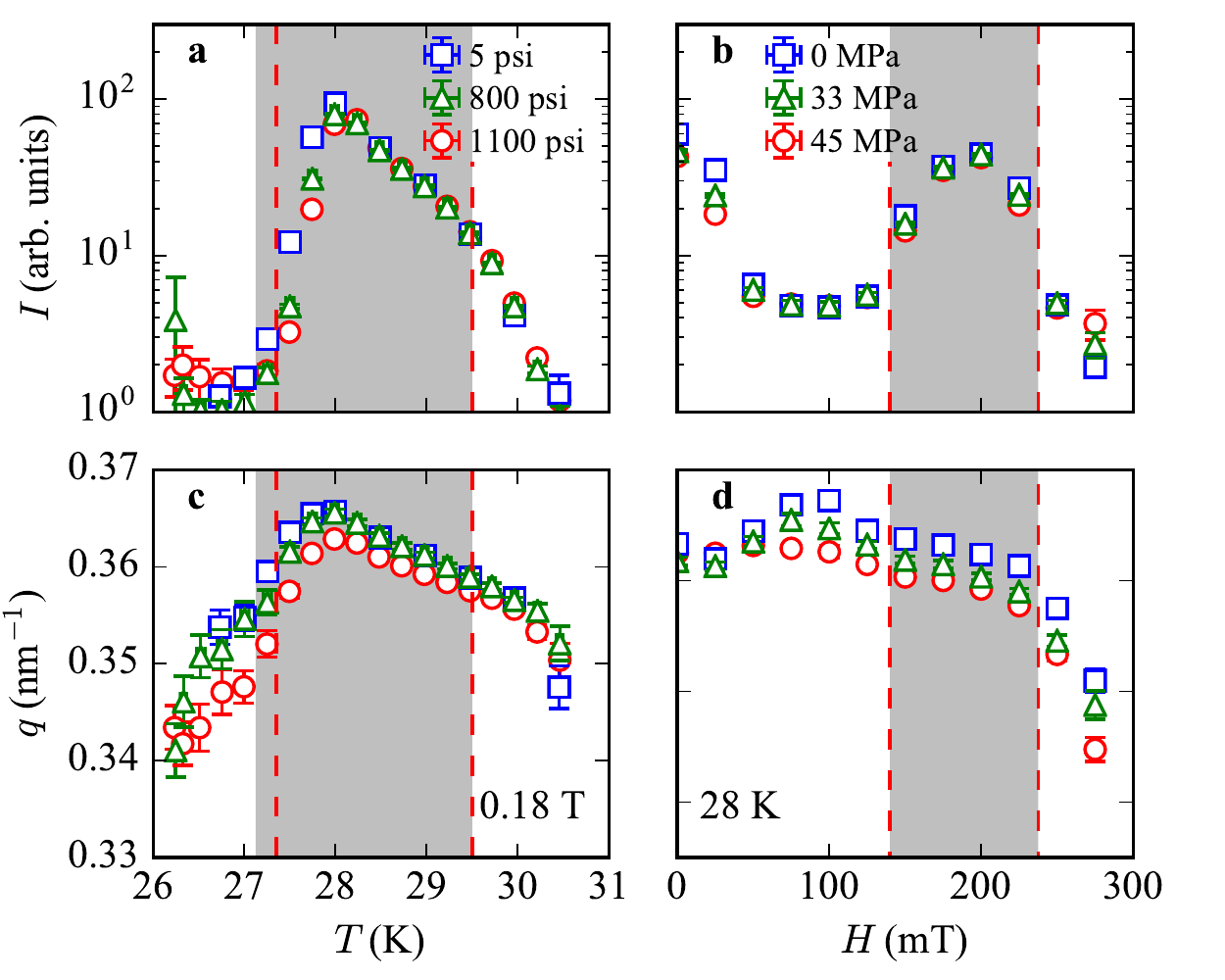}
    \vspace{-0.5cm}
    \caption{Integrated intensity (a and b) and $q$-vector (c and d) as a function of $T$ at 0.18~T (left) and as a function of magnetic field at 28~K (right), for (tensile) 5~psi ($\sigma \approx 0$~MPa) (red circles), 800~psi (33~MPa) (blue squares) and 1100~psi (45~MPa) (green triangles) Approximate equivalent uniaxial pressure values are shown in the legend in panel (b). Horizontal error bars in (a) and (c) represent temperature variance during rocking. All vertical error bars represent errors from least-squares fitting. Shaded regions indicate the skyrmion phase boundaries at ambient pressure, (cf. Fig. 1). Vertical dashed lines show size of skyrmion phase at 1100~psi (45~MPa), obtained from the inflection points in intensity.}
    \label{fig:SANSdata}
\end{center}
\vspace{-1cm}
\end{figure}

Below we describe our experimental results on MnSi with both compressive and tensile strain applied via the strain cell described above. We note that we chose a sample with slightly larger dimensions ($6.5 \times 7 \times 2.25$~mm$^3$) to increase neutron statistics, thus limiting the maximum pressure that could be achieved to only $\sigma\approx 50$~MPa ($\Leftrightarrow\varepsilon$=0.05\%). Furthermore, to test whether the strain response for $\sigma\perp H$ is indeed independent of the exact crystallographic orientation as suggested in Ref.~\onlinecite{Chacon2015}, we have purposefully mis-oriented our sample; the sample was oriented with a [1$\bar{1}$0] axis parallel to, and the [110] and [001] directions perpendicular to, the incident beam (and $H$, respectively, where [110] was $\sim15^{\circ}$ misaligned from the vertical axis ($\parallel \sigma$) (cf. Fig. \ref{fig:detector}(a)).

We chose points in the phase diagram near the lower edge of the A-phase, \textit{i.e.} $H=180$~mT at $T=27$~K and 27.5~K for tensile and compressive strain, respectively, to better illustrate the changes caused by strain. A temperature discrepancy between compressive and tensile strain data may exist due to the use of two different sample environments between the experiments, an older 4.5~T horizontal magnet system, and the newer 11~T system described above. In addition,  the temperature sensor relocated between the two experiments to be closer to the strain cell (before the tensile strain tests). The same sample, described earlier, was used under both strain types.


In Fig. \ref{fig:detector} we present the results of both tensile and compressive strain on MnSi, via SANS detector images obtained by integrating a series of detector images acquired by simultaneously rocking the sample and the magnetic field $\pm10^{\circ}$ about the vertical axis. Background was obtained at 40~K. In Fig. \ref{fig:detector}(a) we see the prototypical skyrmion pattern, consisting of six equidistant peaks on a hexagonal lattice. The relatively weak intensity is due our location in the phase diagram, as mentioned above. Increasing tensile strain to 1000~psi (cf. Fig. \ref{fig:detector}(c)) results in a noticeable decrease in the overall intensity of the pattern, suggesting we are farther from the center of the A-phase than at 200~psi, and thus the A-phase has decreased in size with increased tensile strain. To better illustrate this effect, we show the angle-integrated intensity as a function of $|q|$ between 600~psi and 1100~psi in Fig. \ref{fig:detector}(e); a trend of decreasing intensity with increasing tensile strain is observed, supporting our assertion that the size of the A-phase decreases with increasing tensile strain. In contrast, upon the application of compression we observe a clear increase in intensity between 600~psi and 1100~psi (cf. Fig. \ref{fig:detector}(b) and (d)). There is a unmistakeable jump in intensity around 1100~psi, as seen in Fig. \ref{fig:detector}(f), suggesting compressive strain increases the size of the A-phase, consistent with the previous study of MnSi under compressive strain by A.\ Chacon \textit{et al}.\cite{Chacon2015} Our results on a purposefully mis-oriented sample further demonstrate that compressive strain does not require application along a specific crystallographic direction.

To carefully map the change in the size of the A-phase with the application of tensile strain we performed systematic measurements at $H=0.18$~T, as a function of temperature, and at $T=28$~K as a function of magnetic field, shown as horizontal and vertical lines in Fig. \ref{fig:phase diagram}, respectively. For each temperature and magnetic field a background subtracted detector image was obtained, similar to those shown in Fig. \ref{fig:detector}, and angle-integrated. Angle integrated data was analyzed by least-squares fitting to a Gaussian function. First focusing on the integrated intensity as a function of temperature, seen in Fig. \ref{fig:SANSdata}(a), we can easily identify a trend for the lower temperature boundary to increase in $T$ with increasing pressure, indicating that the overall size of the A-phase decreases in its temperature width. For the maximum applied pressure of about 45 MPa the width of the A-phase decreases 0.3 K. This can be compared to previously reported results where the size of the skyrmion phase was found to decrease when compressive strain $\sigma\parallel H$; the observed decrease of $\approx0.7$~K($\approx 0.5$~K) for of $\sigma$=100 MPa(70 MPa) reported in Ref. \onlinecite{Nii2015}(\onlinecite{Chacon2015}) is commensurate with our result for tensile strain $\sigma\perp H$. This suggests that the magnitude of the response does not depend on the direction the strain is applied in, but is inherently different for the cases where the A-phase is stabilized or destabilized. As a function of magnetic field (cf. Fig. \ref{fig:SANSdata}(b), the phase boundaries do not show any obvious significant shift with increasing pressure, suggesting the A-phase does not change in size along the magnetic field axis. Furthermore, we have observed that the ordering wave vector of the skyrmion lattice increases, non-monotonically overall with increasing tensile strain (cf. Fig. \ref{fig:SANSdata}(c) and (d)) implying that the skyrmion lattice parameter decreases, in agreement with theory.\cite{Lin2015} 

In summary, we report results on the response of the A-phase in MnSi to tensile strain $\sigma$ applied in skyrmion lattice plane, perpendicular to the magnetic field $H$. This result demonstrates that tensile strain destabilizes the skyrmion lattice, opposite to compressive strain for $\sigma\parallel H$ reported in Ref. \onlinecite{Nii2015, Chacon2015}. Our results, combined with previous results under compressive strain,\cite{Nii2015, Chacon2015} suggest that the response to strain is inherently different for the cases where the skyrmion lattice is stabilized or destabilized, where the latter is significantly weaker. Because our experiments where performed with strain purposefully applied in an arbitrary crystallographic direction in the plane perpendicular to $H$, they also demonstrate sufficiently strong magnetic anisotropy is generated regardless of the crystallographic orientation. These results were obtained using a custom developed strain cell optimized for SANS measurements unique in its ability to select between the application of both tensile and compressive strain \textit{in-situ}, which, as our results support, represents a powerful tool for the study of modulated long-period magnetic structures. Finally, we can envision its adaptation to other neutron techniques which require only small sample angle rotations.

\vspace{-0.6cm}
\begin{acknowledgments}
\vspace{-0.3cm}
We thank Shizeng Lin, Cristian Batista and Avadh Saxena for useful discussion, Rex Hjelm, Helmut M. Reiche, and Michael A. Torrez for support during the design of the strain cell, and Kathy Bailey, Doug Armitage, Erik Stringfellow and Jon Smith for technical support during the experiments. Work at Los Alamos National Laboratory (LANL) was performed under the auspices of the U. S. Department of Energy. Research at LANL was funded by the LANL Directed Research and Development program. Research conducted at Oak Ridge National Laboratory's (ORNL) High Flux Isotope Reactor (HFIR) was sponsored by the Scientific User Facilities Division, Office of Basic Energy Sciences, US Department of Energy.
\end{acknowledgments}

\end{document}